\documentstyle[12pt]{article}
\begin{document}

\title{Nonnegative Feynman-Kac Kernels in Schr\"{o}dinger's
Interpolation Problem}

\author{Philippe Blanchard and  Piotr Garbaczewski\thanks{Permanent
address: Institute of Theoretical Physics, University of Wroc{\l}aw,
PL-50 204 Wroc{\l}aw, Poland}\\
Fakult\"{a}t f\"{u}r Physik, Universit\"{a}t Bielefeld,\\
D-33615 Bielefeld, Germany\\
and\\
Robert Olkiewicz\\
Institute of Theoretical Physics, University of Wroc{\l}aw,\\
PL-50 204 Wroc{\l}aw, Poland}

\maketitle
\hspace*{1cm}

\begin{abstract}
The existing   formulations of the Schr\"{o}dinger interpolating
dynamics, which is constrained by the prescribed input-output
statistics data, utilize strictly positive Feynman-Kac kernels.
This implies that the related Markov diffusion processes admit
vanishing probability densities only at the boundaries of the
spatial volume confining the process.

We extend the framework to encompass singular potentials and
associated nonnegative Feynman-Kac-type kernels. It allows to
deal with general nonnegative solutions of the Schr\"{o}dinger
boundary data problem. The resulting stochastic
processes are capable of both developing and destroying nodes
(zeros) of probability densities in the course of their evolution.
\end{abstract}

\newpage

\section{From positive to nonnegative solutions
of para\-bolic equations}

We continue an investigation of the celebrated Schr\"{o}dinger's
boundary data problem, \cite{schr}-\cite{olk2}, of reconstructing
the ''most likely" evolution which interpolates between the
prescribed input-output statistics data (usually analyzed in terms
of nonvanishing probability densities) in a fixed finite-time
interval, interpreted as a duration time of the process.

In the present paper we focus our attention again on stochastic Markov
processes of diffusion-type (see Ref. \cite{olk1} for a jump process
alternative), which are associated with the temporally adjoint pair
of parabolic partial differential equations:
$${\partial _tu(x,t)=\triangle u(x,t)-c(x,t)u(x,t)}\eqno (1)$$
$$\partial _tv(x,t)=-\triangle v(x,t)+c(x,t)v(x,t)$$
Here, $c(x,t)$ is a real function (left unspecified at the moment)
and the solutions $u(x,t)$, $v(x,t)$ are sought for in the time
interval $[0,T]$ under the boundary conditions set at the
time-interval borders:
$${\rho _0(x)=u(x,0)v(x,0)}\eqno (2)$$
$$\rho _T(x)=u(x,T)v(x,T)$$
$$\int_A\rho _0(x)dx=\rho _0(A)\; ,\;
\int_B\rho _T(x)dx=\rho _T(B)$$
We assume that $\rho $ is a probability measure with the density
$\rho (x)$, and $A,B$ stand for arbitrary Borel sets in the
event space. In the above, suitable units were chosen to eliminate
inessential in the present context (dimensional) parameters, and
the process is supposed to live in/on $R^1$.

As emphasized in the previous publications, \cite{blanch}-\cite{olk2},
the key ingredient of the formalism is to specify the function
$c(x,t)$
such that $exp[-\int_0^tH(\tau )d\tau ]$
can be viewed as a strongly continuous semigroup operator
with the generator   $H(t)=-\triangle +c(t)$, associated with the
familiar \cite{simon} Feynman-Kac kernel:
$${(f,exp[-\int_0^tH(\tau )d\tau ]g)=\int dy \int dx
\overline{f}(y)k(y,0,x,t)g(x)=}\eqno (3)$$
$$\int \overline{f}(\omega (0)) g(\omega (t)) exp[-\int_0^tc(\omega
(\tau ),\tau )d\tau ] d\mu _0(\omega )$$
Here $f,g$ are complex functions, $\omega (t)$ denotes a sample path
of the conventional Wiener process and $d\mu _0$ stands for the Wiener
measure. Clearly, the kernel itself  can be explicitly written
in terms
of the conditional Wiener measure $d\mu _{(x,t)}^{(y,s)}$ pinned at
space-time points $(y,s)$ and $(x,t)$, $0\leq s<t\leq T$:
$${k(y,s,x,t)=\int exp[-\int_s^t c(\omega (\tau ),\tau )d\tau ]\: d\mu
^{(y,s)}_{(x,t)}(\omega )}\eqno (4)$$
As long as we do not impose any specific domain restrictions on the
semigroup generator $H(\tau )$, the whole real line $R^1$ is
accessible to the process.
Various choices of the Dirichlet \cite{simon,blanch}
boundary conditions can be accounted for  by the formula (3). If we
replace $R^1$ by  any open subset $\Omega \subset R^1$ with the
boundary $\partial \Omega $,
it  amounts to confining  Wiener sample paths of
relevance to reside in  (be interior to) $\Omega $,
which in turn needs an appropriate measure
$d\mu _{(x,t)}^{(y,s)}(\omega \in \Omega )$ in (4).
This is usually implemented by means of stopping times for the
Wiener process, \cite{nag,blanch,carm,nag1,gol}.

Let $f(x)$ and $g(x)$ be two real functions such that:
$${m_T(x,y)=f(x)k(x,0,y,T)g(y)}\eqno (5)$$
defines a bi-variate density of the probability measure:
$${m_T(A,B)=\int_Adx\int_Bdy\: m_T(x,y)}\eqno (6)$$
i.e. a transition probability of the propagation from the Borel set
$A$ to the Borel set $B$ to be accomplished in the time interval $T$.
In particular, we need the marginal probability densities
 to be defined:
$${\rho_0(x)=m_T(x,\Omega )\; , \; \rho _T(y) =m_T(\Omega ,y)}
\eqno (7)$$
where $\Omega \subset R^1$ is a spatial  area confining  the process.

Formulas (5),(6), can be viewed as special cases  of (3), so
establishing an apparent link between the Schr\"{o}dinger
problem and  the Feynman-Kac  kernels, together with the
related parabolic equations (1).
Assuming that marginal probability  measures (7) and their densities
are given a priori, and a concrete  Feynman-Kac kernel (4)
(with or without Dirichlet  domain restrictions) is specified,
we are within the premises of the Schr\"{o}dinger
boundary data problem.

Let $\overline{\Omega }=\Omega \cup \partial \Omega $ be a closed
subset of $R^1$, or $R^1$ itself. For all Borel sets (in the $\sigma
$-field generated by all open subsets of $\overline{\Omega }$) we
assume to  have known $\rho _0(A)$ and $\rho _T(B)$, hence the
respective densities as well. If the integral kernel $k(x,0,y,T)$
in the expression (5) is chosen to be \it continuous \rm and \it
strictly positive \rm on $\overline{\Omega }$, then the integral
equations (7) can be solved \cite{jam} with respect to the \it
unknown  \rm functions $f(x)$ and $g(y)$.  The solution comprises
two nonzero, locally integrable functions  of the same sign, which
are unique up to a  multiplicative constant.

If, in addition, the kernel $k(y,s,x,t)$, $0\leq s<t\leq T$ is a \it
fundamental solution \rm \cite{fried} of the parabolic system (1) on
$R^1$ (i.e. is a function which solves the forward equation in $(x,t)$
variables, while the backward one in $(y,s)$), then we have defined a
 solution of the system (1) by:
$${u(x,t)\equiv f(x,t)=\int f(y) k(y,0,x,t)dy}\eqno (8)$$
$$v(x,t)\equiv g(x,t)=\int k(x,t,y,T)g(y)dy$$
Moreover, $\rho (x,t)=f(x,t)g(x,t)$ is propagated by the Markovian
transition probability density:
$${p(y,s,x,t)=k(y,s,x,t){{g(x,t)}\over {g(y,s)}}}\eqno (9)$$
$$\rho (x,t)=\int \rho (y,s) p(y,s,x,t) dy $$
$$0\leq s<t\leq T$$
$$\partial _t\rho =\triangle \rho - \nabla (b\rho )$$
$$b=b(x,t)=2{{\nabla g(x,t)}\over g(x,t)}$$
the result, which covers all traditional Smoluchowski diffusions
\cite{blanch,garb}. In that case, $c(x,t)$ is regarded as
time-independent, and the corresponding stochastic process is
homogeneous in time.
The Dirichlet boundary data can be implemented as well,
thus leading to the Smoluchowski diffusion processes with natural
boundaries, \cite{blanch}.
Then, $k(y,s,x,t)$ stands for an appropriate
Green function of the parabolic boundary-data problem,
with the property
to vanish at the boundaries $\partial \Omega $ of $\Omega $.

Recently \cite{olk2}, an extension of the above formalism was
elaborated to
encompass situations when the involved Feynman-Kac kernels are
strictly positive and continuous, but not necessarily
fundamental solutions  of (1) nor even differentiable.
They still give rise to (8) and (9) under suitable regularity
conditions for solutions of the parabolic system (1).
Their existence is not in
conflict with the fact that $k(y,s,x,t)$ itself needs not to be a
solution of any differentiable equation.

Let us also mention that for time-independent potentials,
$c(x,t)=c(x)$ for all $t\in [0,T]$, a number of generalizations is
available \cite{nag,blanch,carm,nag1,gol,combe}-\cite{fuk} to
encompass
the nodal sets of $\rho (x)$ and hence of the associated functions
$f(x),g(x)$.  The drift
$b(x)=\nabla ln\rho (x)={{\nabla \rho (x)}\over {\rho (x)}}$
singularities  do not prohibit the existence of a well defined
Markov diffusion process (9), for which  nodes are unattainable.
In the considered
framework  they are allowed only at the boundaries of the
connected spatial area $\Omega $ confinig the process.

The problem of relaxing the strict positivity (and/or continuity)
demand for Feynman-Kac kernels is nontrivial \cite{fort,beur,jam}
with respect to the eventual construction of the \it unique \rm
Markov process (9). To elucidate the nature of difficulties
underlying
this issue, let us consider  quantally motivated examples of the
parabolic dynamics (1).

\section{Nonlinear parabolic dynamics with the fundamental
solution}

Let us choose the potential function $c(x,t)$ as follows:
$${c(x,t)={x^2\over {2(1+t^2)^2}} - {1\over {1+t^2}}}\eqno (10)$$
for $x\in R^1, t\in [0,T]$.
In view of its local H\"{o}lder continuity
(cf. Ref. \cite{olk2}) with exponent one, and its quadratic
boundedness, the
fundamental solution of the parabolic system  is known to exist
\cite{fried,kal,szyb,bes}. It is constructed via the
 parametrix method \cite{fried}. Among an
infinity of regular solutions of (1) with the potential (10), we can
in particular identify \cite{olk2} solutions of the Schr\"{o}dinger
boundary data problem for the familiar (quantal) evolution:
$${\rho _0(x)=(2\pi )^{-1/2} exp[-{x^2\over 2}] \longrightarrow
\rho (x,t)=[2\pi (1+t^2)]^{-1/2} exp[-{x^2\over {2(1+t^2)}}]}\eqno
(11)$$
They read
$${u(x,t)\equiv f(x,t)=[2\pi (1+t^2)]^{-1/4} exp(-{x^2\over
4}{{1+t}\over {1+t^2}}+ {1\over 2} arctan\: t)}\eqno (12)$$
$$v(x,t)\equiv g(x,t)=[2\pi (1+t^2)]^{-1/4} exp(-{x^2\over 4}{{1-t}
\over {1+t^2}}- {1\over 2} arctan\: t)$$
and, while solving the nonlinear parabolic system (1) (with
$c=\triangle \rho ^{1/2}/\rho ^{1/2}$), in addition they imply the
validity of the Fokker-Planck equation:
$$\rho (x,t)= f(x,t)g(x,t) \rightarrow \partial _t\rho =
\triangle \rho - \nabla (b\rho )$$
$${b(x,t)=2{{\nabla g(x,t)}\over {g(x,t)}} =
- {{1-t}\over {1+t^2}}\: x}\eqno (13)$$
Notice that $p(y,s,x,t)=k(y,s,x,t){{g(x,t)}\over {g(y,s)}}$ is  a
fundamental solution of the first and second Kolmogorov (e.g.
Fokker-Planck) equations in the present case.

Let us recall that a concrete parabolic system corresponding to
solutions (12) looks badly nonlinear.
Our procedure, of first considering
the linear system (but with the potential "belonging" to another,
nonlinear one), and next identifying solutions of interest by means of
the Schr\"{o}dinger boundary data problem,  allows to bypass this
inherent difficulty.
In connection with the  previously mentioned quantal motivation of
ours,
let us define $g=exp(R+S), f=exp(R-S)$  where $R(x,t), S(x,t)$
are real
functions. We immediately realize that (1), (10) provide for
a parabolic alternative to the familiar Schr\"{o}dinger equation
and its temporal adjoint:
$${i\partial _t\psi = -\triangle \psi }\eqno (14)$$
$$i\partial _t\overline{\psi } = \triangle \overline{\psi }$$
with the Madelung factorization $\psi =exp(R+iS)$, $\overline{\psi
}=exp(R-iS)$ involving the previously introduced real functions $R$
and $S$.

\section{Nonlinear parabolic dynamics with unattainable
boundaries: the Green function}

Things seem to be fairly transparent when the parabolic system (1)
allows  for  fundamental solutions. However, even in this case
complications arise if nodes of the probability density are admitted.
The subsequent discussion has a quantal origin  again,
and comes from  the free Schr\"{o}dinger propagation (14) with the
specific choice of the initial data:
$${\psi _0(x)=(2\pi )^{-1/4} \: x\: exp(-{x^2\over 4})
\longrightarrow
}\eqno (15)$$
$$\psi (x,t)=(2\pi )^{-1/4} {x\over {(1+it)^{3/2}}}\: exp[-{x^2\over
{4(1+it)}}]$$
such  that  our nonstationary dynamics example displays a stable
node at $x=0$ for all times .

The  parabolic system (1) in this case involves the potential
function:
$${c(x,t)={{\triangle \rho ^{1/2}(x,t)}\over {\rho ^{1/2}(x,t)}}=
{x^2\over {2(1+t^2)^2}}- {3\over {1+t^2}}}\eqno (16)$$
$$\rho (x,t)=(2\pi )^{-1/2}(1+t^2)^{-3/2}\: x^2\: exp[-{x^2\over
{2(1+t^2)}}]$$
The polar (Madelung) factorization of Schr\"{o}dinger wave functions
implies:
$${R(x,t)=ln\: \rho ^{1/2} (x,t)}\eqno (17)$$
$$x>0\rightarrow S(x,t)=S_+(x,t)=
{x^2\over 4}{t\over {1+t^2}} -{3\over 2}arctan\: t
$$
$$x<0\rightarrow S(x,t)=S_-(x,t)=
{x^2\over 4}{t\over {1+t^2}}-{3\over 2}arctan \: t \:
+\: \pi $$
Although $S(x,t)$ is not defined at $x=0$, we can introduce continuous
functions $f=exp(R-S)$ and $g=exp(R+S)$ by employing the step function
$\epsilon (x)=0$ if $x\geq 0$ and $\epsilon (x)=1$ if $x<0$. Then, the
candidates for solutions of the parabolic system (1) with the
potential (16) would read:
$${v(x,t)\equiv g(x,t)=(2\pi )^{-1/4} (1+t^2)^{-3/4} |x|\:
exp(-{x^2\over 4}{{1-t}\over {1+t^2}})\:
exp[-{3\over 2}arctan\: t + \pi
\epsilon (x)]}\eqno (18)$$
$$u(x,t)\equiv f(x,t)= (2\pi )^{-1/4}(1+t^2)^{-3/4} |x|\:
exp(-{x^2\over
4}{{1+t}\over {1+t^2}})\:
exp[{3\over 2}arctan\: t - \pi \epsilon (x)]$$
For all $x\neq 0$ we can define the forward drift
$${b(x,t)=2{{\nabla g(x,t)}\over {g(x,t)}}={2\over x} - x\: {{1-t}
\over
{1+t^2}}}\eqno (19)$$
which displays a singularity at $x=0$.
Nonetheless, $(b\rho )(x,t)$ is a smooth function and the
Fokker-Planck
equation $\partial _t\rho =\triangle \rho - \nabla (b\rho ) $
holds true
on the whole real line $R^1$, for all  $t\in [0,T]$.
Notice that there is no current through $x=0$, since $v(x,t)=2\nabla
S(x,t)={{xt}\over {1+t^2}}$ vanishes at this point for all times.

Our functions $f(x,t), g(x,t)$ are continuous on $R^1$, which  however
does not imply their differentiability. Indeed, they solve the
parabolic
system (1) with the potential (16) \it not \rm on $R^1$  but on
$(-\infty ,0)\cup (0,+\infty )$. Hence, almost everywhere on $R^1$,
with the exception of $x=0$.

An apparent obstacle arises because of this subtlety: these functions
are \it not \rm even weak solutions of (1), because of:
$${\int_{-\infty }^{+\infty } \partial _tf(x,t)\phi (x) dx +
\int_{-\infty }^{+\infty }\nabla f(x,t) \nabla \phi (x) dx +}\eqno
(20)$$
$${1\over 2} \int_{-\infty }^{+\infty }c(x,t)f(x,t)\phi (x) dx \:
\neq 0
$$
for every test function $\phi $ such that $\phi (0)\neq 0$,
continuous and with support on a chosen compact set (e.g. vanishing
beyond this set).

One more obstacle arises, if we notice that $c(x,t)$, (16) permits the
existence of the unique, bounded and strictly positive fundamental
solution   for the parabolic system (1). Then, while having singled
out
a fundamental solution and the boundary density data
$\rho _0(x), \rho
_T(x)$ consistent with (16), we can address the Schr\"{o}dinger
boundary data problem associated with (2),(3):
$${u(x,0)\int k(x,0,y,T)v(y,T)dy = \rho _0(x)}\eqno (21)$$
$$v(x,T)\int k(y,0,x,T)u(y,0)dy =\rho _T(x)$$
expecting that a unique solution $u(x,0),v(x,T)$ of this system of
equations implies an identification $u(x,0)=f(x,0)$ and
$v(x,T)=g(x,T)$.

However, it is not the case and our $f(x,t),g(x,t)$ do not
come out as
solutions of the Schr\"{o}dinger problem,
if considered on the whole real line $R^1$,
on which the fundamental solution sets rules of the game.

Indeed, let us assume that (21) does hold true if we choose
$u(x,0)=f(x,0),
v(x,T)=g(x,T)$, with $f$ and $g$ defined by (18). Since,
in particular we have
$${g(x,T)\int k(y,0,x,T)f(y,0)dy = g(x,T)f(x,T)}\eqno (22)$$
then for $x\neq 0$ there holds:
$${f(x,T)=\int k(y,0,x,T)f(y,0)dy}\eqno (23)$$
Both sides of the last identity represent continuous functions, hence
the equality is valid point-wise (i.e. for every $x$). We know that
$f(y,0)$ is continuous and bounded on $R^1$, and $k(y,0,x,T)$ is a
fundamental solution of (1). Hence the right-hand-side of (23)
represents a regular solution of the parabolic equation.
Such solutions
have continuous derivatives, while our left-hand-side function
$f(x,T)$ certainly does not share this property.
Consequently, our assumption
leads to a contradiction and (23) is invalid in our case.

It means that the fundamental solution (e.g. the corresponding
Feynman-Kac kernel) associated with (16) is inappropriate for the
Schr\"{o}dinger problem analysis, if the interpolating probability
density is to have nodes (i.e. vanish at some points).

In our case, $x=0$ is a stable node of $\rho (x,t)$, and is a
time-independent repulsive obstacle for the stochastic process. An
apparent way out of the situation comes by considering two
non-communicating processes, which are separated by the unattainable
barrier at $x=0$, \cite{gol,carlen,blanch,alb}.

The function
$${f_+(x,t)=(2\pi )^{-1/4} (1+t^2)^{-3/4}\: x\: exp(-{x^2\over
4}{{1+t}\over {1+t^2}})\: exp({3\over 2}arctan\: t)}\eqno (24)$$
$$x\in [0,\infty ) \;  ,\;  t\in [0,T]$$
is a regular solution \cite{fried} of the first initial-boundary value
problem for $\partial _tf=\triangle f - cf$ specified by:
$${f_+(x,0)=(2\pi )^{-1/4} \: x\: exp(-{x^2\over 4})}\eqno (25)$$
$$f_+(0,t)=0$$
Then, instead of the fundamental solution,
we need to utilize the Green function of the problem.
To distinguish it from the fundamental
Feynman-Kac kernel $k$ we shall denote this Green function $k_+$.
Its existence is granted by the very existence of the fundamental
solution for the considered potential (16), see Ref. \cite{freid}.

The Green function $k_+(y,s,x,t)$ is a unique function such that
for every  $\phi $ continuous on $(0,\infty )$ and with a compact
support, the function:
$${u(x,t)=\int_0^{\infty }k_+(y,s,x,t)\phi (y)dy}\eqno (26)$$
is a solution of $\partial _tu=\triangle u - c(x,t)u$ in
$(0,\infty )\times (s,T)$, with the properties:
$lim_{t\downarrow s}u(x,t)=\phi (x)$  for all $x\in [0,\infty )$, and
 $u(0,t)=0$ for all $t\in (s,T]$.
Moreover, for every $(y,s)\in (0,\infty )\times [0,T)$ the function
$k_+$ is strictly positive in $(0,\infty )\times (s,T)$ and
$k_+(y,s,0,t)=0$ for all $t\in (s,T]$.

The uniqueness of solutions for the first initial-boundary value
problem
implies  the validity of the semigroup composition rule for $k_+$
(Chapman-Kolmogorov identity, which anticipates the
Markov property of the constructed stochastic process).

In  this (uniqueness) connection a more detailed comment is
necessary.

For all $s\in [0,T)$, the function $f_+(x,s)$ can be uniformly
approximated by a sequence of continuous functions $\phi _n^s(x)$
such that, for each natural number $n$ the support of
$\phi _n^s$ is compact in $(0,\infty )$.
There follows that the sequence of solutions of the
first initial-boundary value problem  $\partial _tu=\triangle u - cu$,
given by:
$${u_n(x,t)=\int_0^{\infty } k_+(y,s,x,t)\phi _n^s(y)dy}\eqno (27)$$
is uniformly convergent to the solution $f_+(x,t)$. It implies that
for any $s<t$ we have
$${f_+(x,t)=\int_0^{\infty } k_+(y,s,x,t)f_+(y,s)dy}\eqno (28)$$

Now, let us consider
$${g_+(x,t)=(2\pi )^{-1/4}(1+t^2)^{-3/4}\: x\:
exp(-{x^2\over 4}{{1-t}\over {1+t^2}}\: exp(-{3\over 2}arctan\: t)}
\eqno (29)$$
which is the solution of the first initial-boundary value problem for
the adjoint parabolic equation
$${\partial _tv=-\triangle v + cv}\eqno (30)$$
$$g_+(x,0)=(2\pi )^{-1/4}\: x\: exp(-{x^2\over 4})$$
$$g_+(0,t)=0$$
for all $t\in [0,T]$. Let $k_+^*$ denotes the Green function of this
adjoint equation. For every continuous function $\phi $ with a compact
support in $(0,\infty )$  the formula
$${v(y,s)=\int_0^{\infty } k_+^*(x,t,y,s)\phi (x)dx}\eqno (31)$$
with $s<t$, defines the solution of the first
initial-boundary  problem for
the adjoint  equation. The previous arguments (at least for $T<1$,
modulo appropriate rescalings) apply in this case as well. We conclude
that  there holds
$${g_+(y,s)=\int_0^{\infty }k_+^*(x,t,y,s)g_+(x,t)dx}\eqno (32)$$
But, we have  \cite{fried}:
$${k_+^*(x,t,y,s)=k_+(y,s,x,t)}\eqno (33)$$
for all $x,y\in (0,\infty )$,  and $k_+^*(x,t,0,s)=0$.
So, we can write:
$${g_+(y,s)=\int_0^{\infty } k_+(y,s,x,t)g_+(x,t)dx}\eqno (34)$$
for $y>0$, while $g_+(y,s)=0$ if $y=0$.

All that finally allows  us to introduce the transition
probability density of
the Markov process respecting the stable repulsive boundary at $x=0$
as follows:
$${p_+(y,s,x,t)=k_+(y,s,x,t){{g_+(x,t)}\over {g_+(y,s)}}}\eqno (34)$$
$$y\in (0,\infty )\; , \; x\in [0,\infty )\; ,\; 0\leq s<t\leq T$$
The behaviour of $p_+$ at $y=0$ is to some extent irrelevant, and may
involve a discontinuity. But, an innocent modification on the set of
measure zero is allowed, and we choose  $p(0,s,x,t)=\delta (x)$.

It completes the definition of the transition probability density of
the Markov process, which is consistent with the dynamics of
$\rho (x,t)$. For all $x\in[0,\infty )$, we have
$\rho (x,t)=\int_0^{\infty }p_+(y,s,x,t)\rho (y,s)dy, \:
0\leq s<t\leq T$
and also $\int_0^{\infty } p_+(y,s,x,t)dx=1$ for all
$y\in [0,\infty )$.

However, in view of $b_+(x,t)=2{{\nabla g_+(x,t)}\over
{g_+(x,t}}={1\over x}-x{{1-t}\over {1+t^2}}$, which is singular at
$x=0$, the density $p_+(y,s,x,t)$ \it cannot \rm by itself be a
Green function for the associated Fokker-Planck equation
$\partial _t\rho =\triangle \rho -\nabla (b\rho )$, if considered on
the whole of $R^1$.
The equation $\partial _tp_+=\triangle p_+ - \nabla (bp_+)$ holds true
 in the open set $(0,\infty)\times (0,T)$.

By combining the known results \cite{carm,gol,combe}-\cite{fuk} about
the unattainability of nodes by the diffusion process on $R_+$
(respectively, on $R_-$), we conclude that $p_+(y,s,x,t)$ ($p_-$,
respectively ) is a transition probability density of the diffusion
with the density $\rho (x,t)$, (16),forward drift
$b(x,t)$, (19) for which $x=0$ is an inaccessible repelling barrier.
It remains in conformity with situations met in the conservative
$c(x,t)=c(x)$ cases, when an (ergodic \cite{alb}) decomposition into
the non-communicating due to nodes processes, is generic.

On both semi-axes, the respective strictly positive, continuous
(domain-restricted Feynman-Kac) kernels are defined almost everywhere,
except for the barrier location $x=0$, where they vanish.
Thus, if considered on $(-\infty ,0]$ or $[0,+\infty )$, the
respective integral kernels are \it non-negative \rm , and no longer
strictly positive.
Moreover, they seem to need to be considered separately on $R_+$ and
$R_-$.

It is instructive to add that the existence of a node at time $t=0$
does
not automatically imply its survival for times $t>0$, and in reverse
(while in the present context of non-negative kernels).

\section{The "Wiener exclusion"}

The conventional definition of the Feynman-Kac kernel (in the
conservative case)
$${exp[-t(-\triangle +c)](y,x)=\int exp[-\int_0^t
c(\omega (\tau ))d\tau ] d\mu _{(x,t)}^{(y,0)} (\omega )}\eqno (35)$$
comprises all sample paths of the Wiener process on $R^1$, providing
merely for their nontrivial  redistribution by means of the
Feynman-Kac weight $exp[-\int_0^tc(\omega (\tau ))d\tau ]$ assigned to
each sample path $\omega (s): \omega (0)=y, \omega (t)=x$.

Assume that $c(x)$ is bounded from below and locally (i.e. on compact
sets) bounded from above. Then, the kernel is strictly positive and
 continuous \cite{simon}.

For $c=0$ we deal with the conditional Wiener measure
$${exp(t\triangle )(y,x)=\mu _{(x,t)}^{(y,0)} [\omega (s)\in R^1;
0\leq s\leq t]=}\eqno (36)$$
$$\mu[\omega (s)\in R^1;\omega (0)=y,
\omega (t)=x; 0\leq s\leq t]$$
pinned at space-time points $(y,0)$ and $(x,t)$.

The previous discussion indicates that $R_-$ is
inaccessible for all sample paths originating from $R_+$.
In reverse, $R_+$ is inaccessible for those from $R_-$.
As well, we may confine the process to an arbitrary closed subset
$\Omega \subset R^1$, or enforce it to avoid ("Wiener exclusion" of
Ref. \cite{simon1}) certain areas in $R^1$.

In this context, it is instructive to know that \cite{simon} for an
arbitrary open set $\Omega $, there holds:
$${exp(t\triangle _{\Omega }) (y,x)= \mu _{(x,t)}^{(y,0)}
[\omega (s)\in
\Omega , 0\leq s\leq t]}\eqno (37)$$
which is at the same time  a  definition of  the operator
$-\triangle _{\Omega }$,
i.e. the Laplacian with Dirichlet boundary conditions, and that of
the associated semigroup kernel. This formula provides us with the
conditional Wiener measure which is \it confined \rm to the interior
of a given open set,\cite{gin,brat,blanch}.

We can introduce an analogous measure, which is confined to the \it
exterior \rm of a given closed subset $S\subset R^1$.
In case of not
too bad sets (like an exterior of an interval  in $R^1$ or a ball in
$R^n$, the corresponding integral kernel in known \cite{simon} to be
positive and continuous. Technically, if $S$ is a (regular) closed set
such that the Lebesgue measure of $\partial S$ is zero, then:
$${exp(t\triangle _{R\backslash S})(y,x)=\mu _{(x,t)}^{(y,0)}
[\omega (s)
\notin S; 0\leq s\leq t]}\eqno (38)$$

The Feynman-Kac spatial redistribution of Brownian paths  can
be extended to cases (37),
(38) through the general formula valid for any $f,g \in L^2(\Omega )$,
where $\Omega $ is any open set of interest (hence $R\backslash S$, in
particular):
$${(f,exp(-tH_{\Omega } )g)=\int_{\Omega } \overline {f}(\omega (0))
g(\omega (t))exp[-\int_0^tc(\omega (\tau ))d\tau ] d\mu _0(\omega )}
\eqno (39)$$
It gives rise to the integral kernel comprising the restricted Wiener
path integration, which is defined at least almost everywhere in
$x,y$.
Then, its continuity is not automatically granted.
We can also utilize
a concept of the first exit time $T_{\Omega }$ for the sample path
started inside $\Omega $ (or outside $S$)
$${T_{\Omega }(\omega )=inf[t>0, X_t(\omega )\notin \Omega ]}\eqno (40)$$
where $X_t$ is the random variable of the process. Then, we can write,
\cite{carm,nag1,blanch}
$${exp(-tH_{\Omega })(y,x)=\int exp[-\int_0^tc(\omega (\tau ))d\tau ]
d\mu _{(x,t)}^{(y,0)} [\omega ; t<T_{\Omega }]=}\eqno (41)$$
$$\int_{\Omega }exp[-\int_0^tc(\omega (\tau ))d\tau ]d\mu
_{(x,t)}^{(y,0)}(\omega )$$
It is an integration restricted to these Brownian paths, which while
originating from $y\in \Omega $ at time $t=0$ are conditioned to
reach $x\in \Omega $ at time $t>0$ without crossing (but possibly
touching) the boundary $\partial S$ of $S$.
The contribution from paths which would touch the boundary without
crossing, for at least one instant $s\in [0,t]$ is of Wiener measure
zero, \cite{gin}.

In case of processes with unattainable boundaries,  with
probability 1, there is no sample path which could possibly
reach the barrier at any instant $s<\infty $.

The above discussion made an implicit use of the integrability
property
$${\int_0^tc(\omega (s))ds <\infty}\eqno (42)$$
for $\omega \in R^1, 0\leq s\leq t$., in which case
the corresponding
integral kernel (for bounded from below potentials) is strictly
positive.
Then, if certain areas are inaccessible to the process, it occurs
excusively \cite{combe}-\cite{fuk} due to the drift singularities,
which are capable of "pushing" the sample paths away
from the barriers.

The previous procedure can be extended to the singular
\cite{faris}-\cite{klaud3} potentials, which are allowed to diverge.
Their study was in part motivated by the so called
Klauder's phenomenon
(and the related issue of the ground state degeneracy of quantal
Hamiltonians), and had received a considerable attention in the
literature.

In principle, if $S$ is a closed set in $R^1$ like before, and
$c(x)<\infty $ for all $x\in \Omega =R\backslash S$,
while $c(x)=\infty
$ for $x\in S$, then depending on how severe the singularity is,
we can
formulate a  criterion to grant the exclusion of certain sample
paths of the process and hence to limit an
availability of certain spatial areas to the random motion.
Namely, in case of (42) nothing specific happens, but if we have
$${\int_0^t c(\omega (\tau ))d\tau =\infty }\eqno (43)$$
for $\omega (\tau ) \in S$ for some $\tau \in [0,t]$, then the
"Wiener exclusion" certainly appears: we are left with
contributions from these sample paths only for which (43) does not
occur.  Unless the respective set is of Wiener measure zero.

The area $\Omega $ comprising  the relevant sample paths is then
selected as follows:
$${\Omega =[\omega ; \int_0^tc(\omega (\tau ))d\tau < \infty ]}\eqno
(44)$$

In particular, the criterion (44) excludes from considerations sample
paths which cross $S$ and so would establish a communication
between the distinct connected components of $\Omega $.

The singular set $S$ can be chosen to be of Lebesgue measure zero and
contain a finite set of points dividing $R$ into a finite number
of open
connected components. With each open and connected subset $\Omega
\subset R^1$ we can \cite{faris} associate a strictly positive
Feynman-Kac kernel, which can be expected to display continuity.

Since the respective potentials diverge on $S$, their behaviour in a
close neighbourhood of nodes is quite indicative. For, if $\omega _S$
is a Wiener process sample path which is bound to cross  a node at
$0\leq s\leq t$, then the
corresponding contribution to the path integral vanishes.  Such paths
are thus excluded from consideration.
If their subset is sizable (of nonzero Wiener measure), then the
eliminated contribution
$${\int_{\omega _{S}} exp[-\int_0^tc(\omega _S(\tau ))d\tau] d\mu
_{(x,t)}^{(y,0)}(\omega )= 0 }\eqno (45)$$
is substantial in the general formula (35).

At the same time, we get involved a nontrivial domain property of
the semigroup generator $H=-\triangle +c$ resulting in the so called
ground state degeneracy \cite{faris,klaud,klaud1}.
Let us recall (Theorem 25.15 in Ref. \cite{simon})  that if $c$ is
bounded from below and locally bounded from above, then the ground
state function of $H=-\triangle +c$ is everywhere strictly positive
and thus bounded away from zero on every compact set.

\section{Singular potentials and the ground state degeneracy}

Our further discussion will concentrate mainly on singular
perturbations
of the harmonic potential. Therefore, some basic features of the
respective parabolic problem are worth invoking.
The eigenvalue problem (the temporally adjoint parabolic system
now trivializes):
$${-\triangle g+(x^2-E)g=0=\triangle f-(x^2-E)f}\eqno (46)$$
has well known solutions labeled by $E_n=2n+1$ with $n=0,1,2,...$.
In particular, $g_0(x)=f_0(x)={1\over {\pi ^{1/4}}}exp(-{x^2\over 2})$
is the unique nondegenerate ground state solution. The corresponding
Feynman-Kac kernel reads
$$exp(-tH)(y,x)=k(y,0,x,t)=k_t(y,x)=$$
$${(\pi )^{-1/2} (1-exp(-t))^{-1/2}
exp[-{{x^2-y^2}\over 2} - {{(yexp(-t) -x)^2}\over 2}]}\eqno (47)$$
$$\partial _tk= -\triangle _xk + (x^2-1)k$$
and the invariant probability density $\rho (x)=f(x)g(x)=(\pi
)^{-1/2}exp(-x^2)$ is preserved in the course of the time-homogeneous
diffusion process with the transition probability density
$${p(y,s,x,t)=k_{t-s}(y,x) {{g(x)}\over {g(y)}}}\eqno (48)$$
We have $p(y,s,x,t)=p(y,0,x,t-s)$.

Notice the necessity of the eigenvalue correction of the potential
both
in (35) and (36), which is indispensable to reconcile the functional
form of the forward drift $b(x)=2\nabla ln g(x)=-2x$ with the general
expression for the corresponding (to the diffusion process) parabolic
system potential
$${c=c(x,t)= \partial _t ln\: g + {1\over 2} ({b^2\over 2} + \nabla
b)}\eqno (49)$$
which equals $c(x)=x^2-1$ in our case.

Let us pass to the  singular (degenerate) problems.
\\

{\bf Example 1}:
The potential \cite{nag}:
$${c(x)=x^2+{{\gamma ^2}\over {x^{2(\gamma +1)}}} - {{\gamma (\gamma
+1)}\over {|x|^{2+\gamma }}} - {{2\gamma }\over {|x|^{\gamma }}}}\eqno
(50)$$
with $\gamma >0, x\in R^1$, is singular at $x=0$ and is a
well defined even
function if otherwise. We can give \cite{nag} a solution to the
stationary parabolic system
$${-\triangle g+(c-1)g=0=\triangle f - (c-1)f}\eqno (51)$$
in terms of
$${g(x)=f(x)=exp[-({1\over {|x|^{\gamma }}} + {x^2\over 2})]}\eqno
(52)$$
In this case, an invariant density  up to normalization reads $\rho
(x)=(fg)(x)=g^2(x)$ and is integrable on $R^1$. It vanishes at $x=0$
and at both spatial infinities.

By independent arguments \cite{gol}-\cite{fuk} we know that a Markov
diffusion process  preserving $\rho (x)$ can be consistently defined.
The node $x=0$ is unattainable in view of the appropriate
singularity of the forward drift:
$${b(x)=2\nabla ln\: g(x)= sgn\: x\: {{2\gamma }
\over {|x|^{1+\gamma }}}
- 2x}\eqno (53)$$
which pushes sample paths away from the node. Hence, there is no
communication (realised by sample paths of the process) between $R_+$
and $R_-$. Like in case of $n>1$ eigenfunctions of the harmonic
oscillator, we deal with the totally disjoint (ergodic, \cite{alb})
components of the would-be global \cite{carlen} diffusion.

This feature is nicely manifested in  the apparent domain degeneracy
of the associated semigroup generator $H=-\triangle +(c-1)$.
Namely, $Hg=0$
is simply an eigenvalue problem. Let us define
$g_+(x,t)=g(x,t)$ for $x>0$ and $g_+(x,t)=0$ for $x\leq 0$, and
 $g_-(x,t)=0$ for $x\geq 0$ while $g_-(x,t)=g(x,t)$ for
$x<0$. The same procedure can be repeated for $f\rightarrow f_{\pm }$.

The functions $g_+$ and $g_-$ ($f_+$ and $f_-$ respectively) belong to
$L^2(R^1)$, and are orthogonal on $R^1$, while corresponding to the
same eigenvalue. It is an obvious  spectral degeneracy of the
respective generator $H$.
As mentioned before, semigroups $exp(-tH)$ with strictly positive
kernels do not have \cite{faris} generators with  the ground state
degeneracy.
On the other hand, if $S$ is the (singular) set of Lebesgue  measure
zero and $R\backslash S$ has $m$ or more connected  components, then
there always exists \cite{faris} a positive $c(x)$ in
$L^1_{loc}(R\backslash S)$ such that the ground state of
$H=-\triangle +c$ is $m$-fold degenerate.

This phenomenon we encounter in connection with (50). Recall that
$L^1_{loc}$ comprises equivalence classes of functions which are
integrable on compact sets (e.g. locally).
\\

{\bf Example 2}:
The canonical (in the context of Refs. \cite{faris}-\cite{cal})
choice of the centrifugal potential:
$${c_E(x)=x^2+{{2\gamma }\over x^2} - E}\eqno (54)$$
generates a well known spectral solution \cite{par,cal}
for $Hg=[-\triangle +c_E(x)]g$.  The eigenvalues:
$${E_n=4n+2+(1+8\gamma )^{1/2}}\eqno (55)$$
with $n=0,1,2,...$ and $\gamma >-{1\over 8}\Rightarrow (1+8\gamma
)^{1/2}>\sqrt {2}$, are associated with the eigenfunctions
of the form:
$${g_n(x)=x^{(2\gamma +1)/2}\: exp(-{x^2\over 2})\: L_n^{\alpha
}(x^2)}\eqno (56)$$
$$\alpha =(1+8\gamma )^{1/2}$$
$$L_n^{\alpha }(x^2)=\sum_{\nu =0}^n {{(n+\alpha )!}\over {(n-\nu )!
(\alpha +\nu )!}}\: {{(-x^2)^{\nu }}\over {\nu !}} \longrightarrow $$
$$
L_0^{\alpha }(x^2)=1\; , \; L_1^{\alpha }(x^2)=-x^2+\alpha +1$$
It demonstrates an apparent double degeneracy of both the ground state
and of the whole eigenspace of the generator $H$. The singularity at
$x=0$ does not prevent the definition of $H=-\triangle + x^2 +
{{2\gamma
}\over {x^2}}$ since  this operator is densely defined on an
appropriate subspace of
$L^2(R^1)$. This singularity is sufficiently severe
to
decouple $(-\infty ,0)$ from $(0,\infty )$ so that
$L^2(-\infty ,0)$ and
$L^2(0,\infty )$ are the invariant subspaces of $H$ with the resulting
overall double degeneracy.
\\

Potentials of the form, \cite{simon,simon1,faris}:
$${c(x)=x^2 + [dist(x,\partial \Omega )]^{-3}}\eqno (57)$$
where $\partial \Omega $ can be identified with $\partial S$, and $S$
is a closed subset in $R^1$ of any (zero or nonzero) Lebesgue measure,
have properties generic to the Klauder's phenomenon.
Because the Wiener
paths are known to be H\"{o}lder continuous of any order ${1\over
2}-\epsilon , \epsilon >0$ and of order ${1\over 3}$ in particular,
there holds $\int _0^tc(\omega (\tau ))d\tau =\infty $ if
$\omega (\tau )\in S$ for some $\tau $.
Conversely, $\int _0^c(\omega (\tau ))d\tau
<\infty $ if $\omega $ never hits $S$.
This implies that the relevant contributions to:
$${(f,exp[-t(-\triangle +c)]g)=\int \overline {f}(\omega (0))
g(\omega
(t)) exp[-\int_0^t c(\omega (\tau ))d\tau ] d\mu _0(\omega )}\eqno
(58)$$
come only from the subset of paths defined by
$${Q_t= [\omega ;\; \int_0^t c(\omega (\tau ))d\tau < \infty
]}\eqno (59)$$
The above argument might seem inapplicable to the centrifugal problem.
However it is not so.  In the discussion of the divergence of certain
integrals of the Wiener process, in the context of Klauder's
phenomenon, it has been proven \cite{klaud2} that for almost
every path from $x=1$
to $x=-1$ (crossing the singularity point $x=0$) there holds
$\int_{t-\delta }^{t+\delta }|\omega (\tau )|^{-1} d\tau = \infty $
for any $\delta >0$.

To be more explicit:
if $\tau _1=\tau _1(\omega )$ is the first time such that the Wiener
process $W(t)=W(t,\omega )$ attains the level (location on $R^1$)
 $W(\tau _1)=1$, then the integral over any right-hand-side
neighbourhood $(\tau _1,\tau _1+\delta )$ of $\tau _1$ diverges:
$${\int_{\tau _1}^{\tau _1+\delta } c(\omega (t)-1)dt=\infty }\eqno
(60)$$
if $\int_{-1}^{+1} c(x)dx = \infty $.\\
In case of the left-hand-neighbourhood of $\tau _1$, we have
$${\int_{\tau _1-\delta }^{\tau _1}c(\omega (t)-1)dt =\infty }\eqno
(61)$$
if $\int_{-1}^0 x c(x)dx =\infty $.

All that holds true in case of the centrifugal potential, thus proving
that the only subset of sample paths, which matters in (58) is (59).
Obviously, $Q_t$ does not include neither paths crossing $x=0$ nor
those which might hit (touch) $x=0$ at any instant.
The singularity is sufficiently severe to create an unattainable
repulsive boundary for all possible processes, which we can associate
with the spectral solution (55),(56).

\section{The singular potential in action}

After the previous analysis one might be left with an  impression that
the appearence of the stable barrier at $x=0$ persisting for all $t\in
[0,T]$, is a consequence of the initial data choice
$\psi _0(0)=0 $  for the involved quantum Schr\"{o}dinger picture
dynamics.  In general it is not so.
For example, $\psi _0(x)=x^2exp(-x^2/4)$ which vanishes at $x=0$, does
not vanish anymore for times $t>0$ of the free evolution.

On the other hand, somewhat surprisingly from the parabolic
(intuition) viewpoint, the node can be dynamically developed from the
nonvanishing initial data and lead to the nonvanishing terminal
data.

Let us consider a complex function:
$${\psi (x,t)=(1+it)^{-1/2} exp[-{x^2\over {4(1+it)}}] \; [{x^2\over
{2(1+it)^2}} +{{it}\over {1+it}}]}\eqno (62)$$
which solves the free Schr\"{o}dinger equation with the initial data
$\psi (x,0)={x^2\over 2}exp(-{x^2\over 4})$. It vanishes at $ x=0$
exclusively at the initial instant $t=0$ of the evolution.

Obviously, there is nothing to prevent us from considering
$${\Psi (x,t)=\psi (x,t-\alpha )}\eqno (63)$$
for $\alpha >0$. It solves the same free equation, but with
nonvanishing initial data. However, the node is developed in the
course
of this evolution at time $t=\alpha $ and instantaneously
desintegrated for times $t>\alpha $.
Here, the Schr\"{o}dinger boundary data problem would obviously
involve
two strictly positive probability densities
$\rho _0(x)=|\Psi (x,0)|^2$
and $\rho _T(x)=|\Psi (x,T)|^2,\; T>\alpha $. It would suggest to
utilize the theory \cite{olk2}, based on strictly positive Feynman-Kac
kernels, to analyze the corresponding interpolating process.
However,  this
tool is certainly inappropriate and cannot reproduce the a priori
known
dynamics, with the node arising at the intermediate time instant.

To handle the issue by means of a parabolic system,
which we can always
associate with a quantum Schr\"{o}dinger picture dynamics, let us
evaluate the potential $c(x,t)$ appropriate for (1).

In view of
$${\rho (x,t)= const \; (1+t^2)^{-5/2} exp[-{x^2\over {2(1+t^2)}}]\;
[{x^4\over 4} - x^2t^2 + t^2(1+t^2)]}\eqno (64)$$
we have  (while setting $w^{1/2}(x,t)=[{x^2\over 4}-x^2t^2 +
t^2(1+t^2)]$):
$${c(x,t)={{\triangle \rho ^{1/2}(x,t)}\over {\rho ^{1/2}(x,t)}}=
{1\over
4} (-{x\over {1+t^2}}+\nabla w)^2+
{1\over 2}(-{1\over {1+t^2}}+ \triangle ln \; w)=}\eqno (65)$$
$${1\over 4}{x^2\over {(1+t^2)^2}}-{1\over 2}{{3x^2-2t^2x}\over
{{x^4\over 4} -t^2x^2+ t^2(1+t^2)}}-{1\over {2(1+t^2)}}+$$
$${1\over 2}{{3x^2-2t^2}\over {{x^4\over 4}-t^2x^2+t^2(1+t^2)}}-
{1\over 4}({{x^3-2t^2x}\over {{x^4\over 4} - t^2x^2 +t^2(1+t^2)}})^2$$
The expression looks desparately discouraging, but its $t\downarrow 0$
(i.e. the initial data ) limit is quite familiar and displays a
centrifugal singularity at $x=0$:
$${c(x,t)={{\triangle \rho ^{1/2}(x,0)}\over {\rho
^{1/2}(x,0)}}={x^2\over 4}+ {2\over x^2} - {5\over 2}}\eqno (66)$$

Since the original, dimensional expression for the centrifugal
eigenvalue problem is \cite{par}:
$${(-{1\over 2}\triangle +{m^2\over 2}x^2 + {\gamma \over x^2})g=
Eg}\eqno (67)$$
$$E_n=m[2n+1+{1\over 2}(1+8\gamma )^{1/2}]$$
with $n=0,1,...$, an obvious adjustment of constants $m=1/2,
\gamma =1$
allows to identify $E=5/2$ as the $n=0$ eigenvalue of the centrifugal
Hamiltonian $H=-\triangle +{x^2\over 4} +{2\over x^2}$.

A  peculiarity of the considered example is that it enables us to
achieve an explicit insight into   an emergence of the
centrifugal singularity and its
subsequent destruction (decay) for times $t>\alpha $, due to the free
quantum evolution.

In view of the degeneracy of the ground-state eigenfunction
${x^2\over 2}exp(-{x^2\over 4})$ of the centrifugal Hamiltonian,
we deal here with
the gradually decreasing communication between $R_+$ and $R_-$, which
results in the emergence of the completely separated (disjoint) sets
$(-\infty ,0)$ and $(0,+\infty )$ at $t=\alpha $,  followed by the
gradual increase of the communiaction  for times $t>\alpha $. By
"communication" we understand that the set of sample paths crossing
$x=0$ forms a subset of nonzero Wiener measure.

It also involves a generalisation (cf. also Refs.
\cite{olk2,klaud3,freid}) to  time-dependent Feynman-Kac kernels:
$${(f,exp[-\int_s^tH(\tau )d\tau ]g)=\int \overline {f}(\omega
(s))g(\omega (t))exp[-\int_s^tc(\omega (\tau ),\tau )d\tau ]\; d\mu
_0(\omega )}\eqno (68)$$
$$Q_{s,t} =[\omega ;\: \int_s^t c(\omega (\tau ),\tau )
d\tau <\infty ]$$
The finiteness condition $\int_s^tc(\omega (\tau ),\tau )d\tau
<\infty
$, surely does not hold true, \cite{klaud2}, if $\delta >0$ is
sufficiently small, cf. (60),(61).

Let us mention that a few  interesting
mathematical questions are  left aside in the present paper.
They deserve a separate, thorough  study
 on their own. For example, even in case of conventional
Feynman-Kac kernels, the weakest possible  criterions allowing for
their continuity in spatial variables  are not yet established.
An issue of
the continuity of the kernel in case of general singular potentials,
needs an investigation as well.

\section{Nonnegative solutions of parabolic equations and the
Schr\"{o}dinger boundary data problem according to R. Fortet}

As emphasized before, one of  motivations for our analysis
was the quantal observation (exploiting the Born statistical
interpretation postulate as the principal building block
of the theory)
that the temporally adjoint pair of Schr\"{o}dinger equations:
$${i\partial _t\psi (x,t)=-D\triangle \psi (x,t) + {1\over
{2mD}}V(x)\psi (x,t)}\eqno (69)$$
$$i\partial _t\overline {\psi }(x,t)=D\triangle \overline {\psi }(x,t)
- {1\over {2mD}}V(x)\overline {\psi }(x,t)$$
by means of the polar decomposition $\psi =exp(R+iS),\overline {\psi
}=exp(R-iS)$ can be transformed into the (hopelessly looking at the
first glance) nonlinearly coupled  parabolic system of the form (1):
$${\partial _t\theta _*=D\triangle \theta _* -{1\over {2mD}}(2Q-V)
\theta
_*}\eqno (70)$$
$$\partial _t\theta =-D\triangle \theta + {1\over {2mD}}(2Q-V)
\theta $$
$$Q=2mD^2{{\triangle \rho ^{1/2}}\over {\rho ^{1/2}}}$$
$$\rho (x,t)=\theta _*(x,t)\theta (x,t)=\overline {\psi }(x,t)\psi
(x,t)$$
for real functions $\theta =exp(R+S),\theta _*=exp(R-S)$. In the above
$\hbar /2m=D$  can be set to restore the traditional notation.

While searching for a probabilistic meaning of the system (70), we had
in fact assumed (see also Refs. \cite{olk,olk2}) to have in hands a
solution of (69), so that $\rho (x,t)$ was known.
Our next step amounts
to replacing the nonlinear parabolic system (70)  by a linear one (1),
(with $D=m=1$) where for each given functional choice of
$c(x,t)={1\over
{2mD}}(2Q-V)$, \it all \rm allowed solutions $u(x,t),v(x,t)$ were
sought
for, including the fundamental one. At this point the crucial role of
the respective Feynman-Kac kernel was disclosed.
Effectively, we have invoked the Schr\"{o}dinger boundary data problem
to pick up the unique  solution from among an infinity of others,
such that the arising probability measure dynamics (if any) is
consistent with the a priori known probability  density  boundary
data.

In this indirect way, a definite justification
for  the probabilistic significance of the nonlinear parabolic system
has been achieved.
The procedure proves to be consistent, thus allowing to investigate
fundamental solutions, Green functions and other solutions of (70).
Even, if viewed as the nonlinearly coupled parabolic system per se.
See e.g. also Refs. \cite{nag,kean} for a discussion of
the "propagation of chaos"
in a system of interacting (coupled) diffusion processes

If the Feynman-Kac kernels are strictly positive, the respective
solutions are strictly positive as well, except for the boundaries
$\partial \Omega $ of the spatial area confining the process.
Once we admit the \it nonnegative \rm Feynman-Kac kernels, we fall
into another theoretical framework,
this of nonnegative solutions of linear
(and nonlinear) parabolic equations, \cite {aron}.
Since we address the general time-independent potentials,
and as we have
seen the time-dependent  domain properties of semigroup generators are
involved, we touch upon an almost undeveloped theory of
construction of
Markov processes associated  with time-dependent Dirichlet forms and
spaces, \cite{oshi,oshi1}.
The standard theory of forms \cite{fuk} does
not work in case of time-inhomogeneous evolutions.

On the other hand, the Schr\"{o}dinger problem  itself needs an
extension to nonnegative Feynman-Kac kernels. The strongest
(uniqueness
of solution) result was established \cite{jam} for strictly positive
kernel functions, a demand which needs to be relaxed for our purposes.

Hence, our major assumption must be that the kernel is a nonnegative
and
continuous function. One can try to relax the continuity condition as
well. Moreover, it appears that the kernel may not be a function,  and
nonetheless one can expect that the main features of our analysis
would persist.

Beurling \cite{beur} has attempted to relax the strict positivity
condition in more than one spatial dimension, with a partial success
only.
An earlier analysis due to Fortet \cite{fort} is of particular
importance in our one-dimensional context. He addressed an issue of
the existence and uniqueness of nonnegative solutions of
the Schr\"{o}dinger
boundary data problem, under an assumption that the kernel is
continuous  and nonnegative in one spatial dimension.

Fortet's integral equations:
$${\rho _1(x)=f(x)\int_{\Omega _2} k(x,y)g(y)dy}\eqno (71)$$
$$\rho _2(x) = g(y)\int_{\Omega _1} f(x)  k(x,y)  dx$$
where $\Omega _1,\Omega _2$ are finite (or not) intervals in $R^1$,
and
$\int_{\Omega _1}\rho _1(x)dx=\int_{\Omega _2}\rho _2(y)dy >0$,
while
$k(x,y)\geq 0$, $\rho _1(x)\geq 0,\rho _2(y)\geq 0$,  were to be
solved with respect to the unknown functions $f(x),g(y)$,
 defined respectively on $\Omega _1,\Omega _2$.

All functions $k(x,y),\rho _1(x),\rho _2(y)$ are by assumption
real and
measurable (and integrable) for $x\in \Omega _1,y\in \Omega _2$.
There are however the additional assumptions which must be respected:
\\
(i) $k(x,y)$ is continuous, bounded from the above and nonnegative
almost everywhere in $\Omega _1\times \Omega _2$, i.e.
except for a set $S$ of
measure zero comprising both $x's$ and $y's$ in $R^1$, \\
(ii) $\rho _1(x)$ and $\rho _2(y)$ are continuous,\\
(iii) let $\overline {A}$ be a closed interval in $\Omega _1$. For a
nonnegative continuous in $\Omega _2$  functions $g(y)$ we demand that
\it if \rm the integral
$G(x) =\int_{\Omega _2}k(x,y)g(y)dy$ is finite almost everywhere on an
\it open \rm subset of $\Omega _1$ containing $\overline {A}$,
then this
integral is uniformly convergent on $\overline {A}$. Analogously with
respect to $y\in \overline {B}\subset \Omega $, with $f(x)\rightarrow
F(y)$ on $\Omega _2$.

Under these hypotheses, the integral equations (71) admit a unique
solution given in terms of two functions:\\
(1) $f(x)$ which is strictly positive  and continuous almost
everywhere,
except for the set of zeros of $\rho _1(x)$,\\
(2) $g(y)$ which has the \it same \rm  zeros as $f(x)$, is strictly
positive almost everywhere and measurable.

Hence, $g(y)$ is not necessarily
continuous, and one should realise that we must have granted the
existence of  $\nabla
ln\: g(x,t)$ as the drift field.
If the function $g(x)$ is continuous, all that fits to our previous
discussion.
However, even in this case  some additional restrictions on
$k(x,y)$ are
necessary  to guarrantee a differentiability  of $f(x,t)=\int
k(y,0,x,t)f(y)dy$ and $g(x,t)=\int k(x,t,y,T)g(y)dy$, and make them
solutions of the time-adjoint parabolic system once we set
$k(x,y)=k(y,s,x,t), 0\leq s<t\leq T$ and select the appropriate
Feynman-Kac kernel.

In connection with the previous centrifugal example, the above
conditions appear to be too restrictive. Let us therefore invoke
another result due to Fortet. Namely, if the above condition (iii) is
replaced by the demand:
$${\int_{\Omega _2}{{\rho _2(y)}\over {[\int_{\Omega _1}
k(z,y)\rho _1(z)dz]}}\: dy\; <\; \infty }\eqno (72)$$
then a unique nonnegative solution of the integral equations (71)
comprises a continuous function $f(x)$ whose zeros coincide with those
of $\rho _1(x)$. The function $g(y)$ is measurable and has zeros of
$\rho _2(x)$, which are not necessarily in common with those of $\rho
_1$.

This result opens a number of interesting propagation scenarios, and
deserves a careful analysis (with the generalization prospects) in
higher dimensions, cf. also for a related discussion in Refs.
\cite{wal,lof}.
Notice, that our centrifugal example exhibits in its simplest version
an intriguing  feature of Fortet's analysis:
we are capable of producing
probability densities $\rho _1(x)$ (initial) and $\rho _2(y)$
(terminal),
which have  \it non-coinciding \rm sets of zeros. The inquiry into
the corresponding Schr\"{o}dinger's interpolating dynamics
is quite an appealing problem.\\

{\bf Acknowledgement}: Both authors receive  a financial support from
the KBN research grant No 2 P302 057 07.

\end{document}